\documentclass[a4paper]{jpconf}
\usepackage{graphicx,url,amssymb,amsmath}

\newcommand{\td}[3]{\frac{d^{#3} #1}{d {#2}^{#3}}} 

\begin{document}
\title{Probing quantum gravity at low energies}

\author{Justine Tarrant, Geoff Beck, and Sergio Colafranesco}

\address{School of Physics, University of the Witwatersrand, Private Bag 3, WITS-2050, Johannesburg, South Africa}

\ead{justine.tarrant@wits.ac.za, geoffrey.beck@wits.ac.za, sergio.colafrancesco@wits.ac.za}

\begin{abstract}
Planck stars form when a collapsing shell of matter within a black hole reaches the Planck density, roughly equivalent to the mass being compressed into a volumetric size near that of the proton, and rebounds outwards. These planck stars have been considered as accounting for both fast radio bursts and short gamma ray bursts, whilst offering a comparatively low energy perspective onto quantum gravity. The observation of such an event would require black hole masses much smaller than a solar mass, which could be provided by primordial black hole dark matter models. We discuss the low energy isotropic background emissions produced by decaying primordial black holes at all epochs and derive constraints from the spectrum of the extragalactic background light. We find that, in order to avoid exceeding known extragalactic background light emissions, we must restrict the total energy emitted at low frequencies by a planck star exploding in the present epoch to be less than $10^{13}$ erg or restrict the primordial black hole population far below any existing limits. This casts doubt on whether exploding planck stars could actually account for fast radio bursts, as they are speculated to in the literature. 
\end{abstract}

\section{Introduction}
The concept of a ``planck star" has been proposed in the literature~\cite{rovelli2014} as producing both non-singular black hole spacetimes and solving the information-loss problem~\cite{hawking1975,hawking1976}. It is argued, through loop quantum gravity calculations~\cite{haggard2014}, that the collapsing shell of matter within the event horizon will stop collapsing at a finite density (the Planck density) due to quantum-gravitational pressure~\cite{rovelli2014} and then proceeds to bounce outward in a manner similar to certain collapsing universes in loop quantum cosmology~\cite{ashtekar2006}. The now expanding matter within the black hole then tunnels through the horizon producing a white hole~\cite{barrau2014b}. The bounce and tunnelling happen very rapidly in the rest-frame of the black hole~\cite{rovelli2014}, but due to time dilation the process takes longer than the lifetime of the universe for black holes above $10^{-7}$ M$_{\odot}$~\cite{barrau2014,barrau2014b} but remains more rapid than Hawking evaporation~\cite{rovelli2014}. The photon gas, captured during Primordial Black Hole (PBH) formation in the early universe, and then released by the white hole retains a temperature on the order of a TeV~\cite{rovelli2017}. This intriguing possibility has been argued~\cite{barrau2014b} to be able to account for both fast radio bursts~\cite{lorimer2007}, with dimensional arguments for emissions around a wavelength equal to the black hole diameter~\cite{rovelli2017}, and short gamma-ray bursts provided by the escaping photon gas~\cite{rovelli2017}. Moreover, this scenario offers a comparatively low energy window into the physics of quantum gravity. However, observing such an event requires the presence of black holes with masses considerably below that of the sun.

Fortunately, the possibility of the formation of low-mass PBHs in the very early universe has a long history in the literature (an early argument can be found in \cite{pbh1}) and have often been put forward as potential candidates for dark matter~\cite{pbh2}. Early interest was focussed upon PBHs with masses below that of the sun, but, stringent constraints (see \cite{capela2013} and references therein) prevent PBHs from composing a large fraction of cosmologically relevant dark matter. These constraints, however, assumed the mass distribution of PBHs to be monochromatic and recent work has been able to show that extended mass distributions can either evade~\cite{kuhnel2017} or satisfy~\cite{bellomo2017} the constraints that ruled out PBHs as a major component of cosmological dark matter. These works focussed on lognormal differential mass distributions, as these have been shown to approximately model the seeding of PBH density perturbations by many different inflationary scenarios~\cite{inf1,inf2,inf3}. In particular, the distributions satisfying existing constraints and supplying the bulk of cosmological dark matter have a mean mass around $10$ M$_{\odot}$~\cite{bellomo2017} which has been part of a new interest in more massive PBHs~\cite{mpbh1,mpbh3,carr2018} as it correlates with the mass range of black holes observed by LIGO~\cite{mpbh2}. 

In this work we will study the possibility of an isotropic radio background produced by exploding planck stars using lognormal mass distributions that evade the existing constraints (with a mean mass around $10^{-9}$ M$_{\odot}$~\cite{kuhnel2017}) and those that satisfy constraints~\cite{bellomo2017} (with a mean mass in the LIGO range). This will be done by comparing predicted background spectra to those of known isotropic backgrounds in the same frequency range, known as the Extra-galactic Background Light (EBL) with data taken from \cite{ebl2017} and references therein. We use this comparison to determine a limit on the maximum fraction of the mass-energy of the PBH that is converted into radio emissions as it decays through the planck star mechanism. In particular we show that, for both PBH mass distributions studied, the maximal energy in radio emissions is around $10^{13}$ erg (at a $3\sigma$ confidence interval) for a $10^{-7}$ M$_{\odot}$ PBH exploding in the present epoch. This suggests that such planck star explosions cannot act as a candidate for fast radio burst emission. These are the first stringent empirical constraints placed on this quantum gravity scenario for black hole decay. 

This paper is structured as follows: in section~\ref{sec:emm} we lay out the formalism for calculating the low-frequency background produced by decaying PBHs up until the present epoch. In section~\ref{sec:ebl} we detail the EBL data used and how we draw constraints from it. Finally, in section~\ref{sec:res} and \ref{sec:conc} we display and discuss our results.

\section{Low-Frequency Emission}
\label{sec:emm}

The flux of the low-frequency background radiation generated by exploding planck stars from the time of matter radiation equality will be taken to be
\begin{equation}
\Phi (\lambda) = \int_{z_{eq}}^{0} dz \, \frac{dV}{dz} N_{\tau} (z) \frac{d\tau}{dz}\frac{c}{2 R_{BH}(z)}\frac{\chi(\lambda,z)}{4 \pi D_L^2} \; ,
\end{equation}
where $\frac{dV}{dz}$ is the differential co-moving volume element, $\frac{2 R_{BH}(z)}{c}$ is the assumed time taken for emission of the energy of the PBH exploding at $z$, $N_{\tau} (z)$ is the differential density of exploding planck stars per unit time at redshift $z$, $\chi(\lambda,z)$ is the spectral energy distribution of an individual explosion and $D_L$ is the co-moving distance at redshift $z$. 

$N_{\tau}$ we calculate as follows: take some differential density of primordial black holes per unit mass $N(M)$ and
\begin{equation}
N(M)dM = N(\tau)d\tau \; .
\end{equation}
Thus, using the relation for the lifetime $\tau$ of the black hole of mass $M$~\cite{rovelli2014}
\begin{equation}
\tau = \left(\frac{M}{M_{pl}}\right)^2 t_{pl} \; ,
\end{equation}
we can find the necessary factor of $\td{M}{\tau}{}$. Here $t_{pl}$ and $M_{pl}$ are the planck time and mass respectively. The form of $N(M)$ will be taken, following both \cite{bellomo2017,kuhnel2017}, to be lognormal, as this has been shown to fit a wide range of inflationary formation models for PBHs~\cite{inf1,inf2,inf3}. $N(M)$ will then be normalised to some fraction $f_{pbh}$ of the total dark matter density being composed of primordial black holes.

At present the functional shape of $\chi$ is unknown for low frequencies~\cite{barrau2014,rovelli2017}. However, it is argued in \cite{barrau2014,rovelli2014,barrau2014b} that the majority of low-frequency emission will take place at a wavelength equal to twice the radius of the exploding black hole $R_{bh}$. Therefore, we will assume the function $\chi$ has a thermal shape and peaks at $\lambda = 2 R_{bh}$. The energy emitted in radio will then be normalised to some fraction $\chi_0$ of the total mass-energy of the black hole. Additionally, cosmological redshift effects must be taken into account, we do so with the formula~\cite{barrau2014b} 
\begin{equation}
\lambda_{obs} \sim \lambda_{em} (1+z) \sqrt{\frac{H_0^{-1}}{6 k \Omega_{\Lambda}^{1/2}} \sinh^{-1}{\sqrt{\frac{\Omega_{\Lambda}}{\Omega_m (1+z)^3}}}} \; ,
\end{equation}
where $\lambda_{obs}$ and $\lambda_{em}$ are the observed and emitted wavelengths, $H_0$ is the Hubble constant from \cite{planck2014}, $\Omega_m$ and $\Omega_{\Lambda}$ are the matter and cosmological constant density parameters (also from \cite{planck2014}), and $k = 0.05$ is a pure number originating from loop quantum gravity calculations~\cite{haggard2014}.

\section{EBL Data}
\label{sec:ebl}
We source EBL spectral data from \cite{ebl2017} and references therein. This will then be compared to the projected low-frequency background emission from planck star explosions. This comparison will be informative as we should expect the planck star emissions over all epochs to contribute an extra component to isotropic low-frequency backgrounds. Any value of the free parameter product $f_{pbh} \chi_0$ that allows the planck star spectrum to exceed the known background by $3\sigma$ confidence level or more is then taken as excluded. 

\section{Results}
\label{sec:res}
 
\begin{figure}[htbp]
	\centering
	\resizebox{0.7\hsize}{!}{\includegraphics{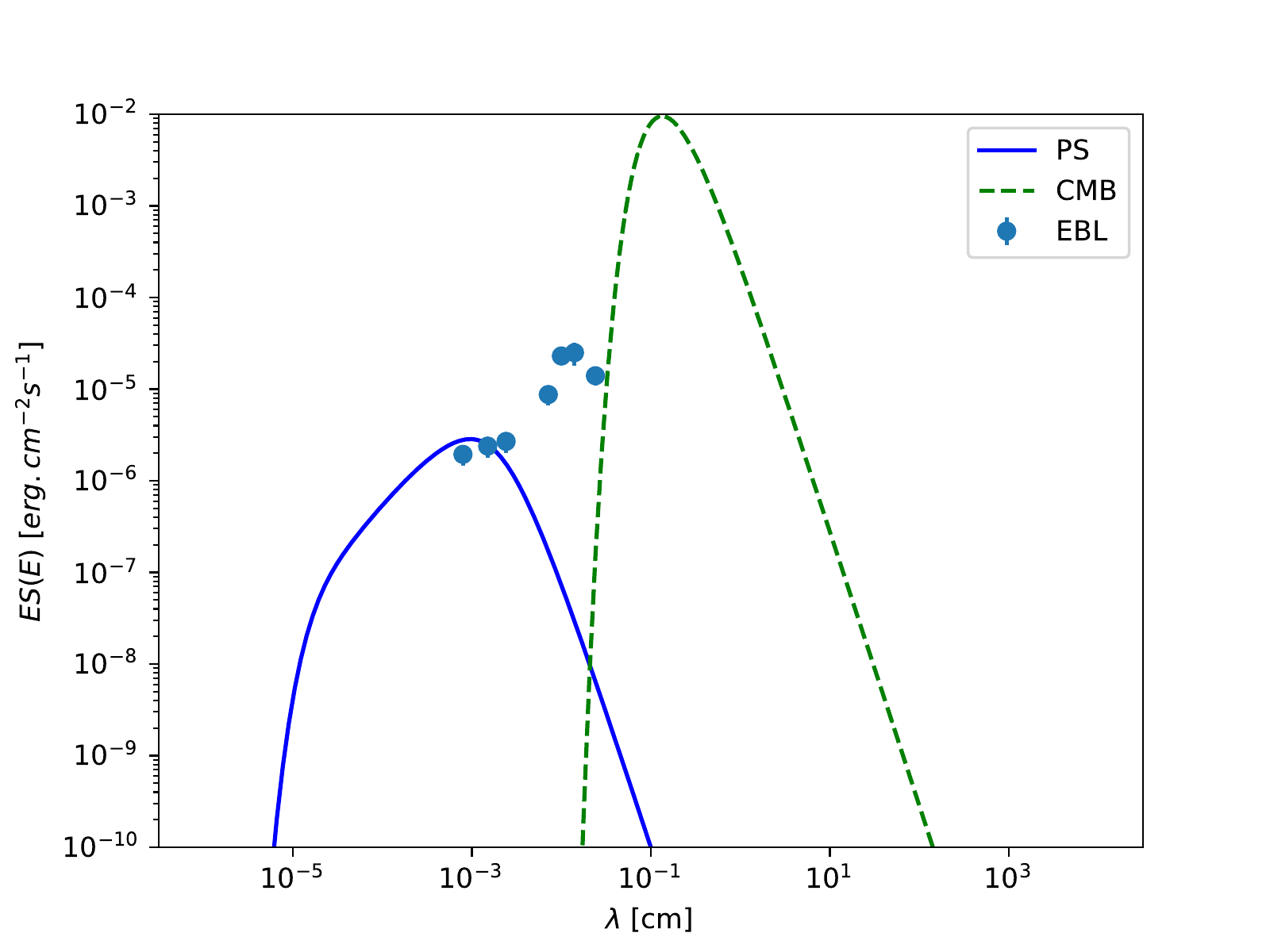}}
	\caption{Comparison of low-frequency planck star background (PS) to EBL data from \cite{ebl2017} and the CMB spectrum~\cite{planck2014}. Values of $\chi_0 f_{pbh} > 10^{-34}$ are excluded at a $3\sigma$ confidence level.}
	\label{fig:ps1}
\end{figure}

In figure~\ref{fig:ps1} we see an example where we take a mass function $N(M)$ from \cite{bellomo2017} with mean $\mu = 10.0$ M$_{\odot}$ and deviation $\sigma = 0.25$ (as this can have PBHs constitute almost all dark matter when considering limits from the CMB, micro-lensing, and ultra-faint dwarf galaxies). The minimal $\chi^2$ fitting performed yields a result that $\chi_0 f_{pbh} < 10^{-34}$. Since for this model of the PBH mass function $f \sim 1$~\cite{bellomo2017}, we can see that this yields an extremely stringent limit on the amount of energy emitted in low-frequency by these planck star explosions. If we take a planck star, exploding at the present epoch, with mass $\sim 10^{-7}$ M$_{\odot}$ the total energy yield of the explosion is $\sim 10^{47}$ erg, making it an attractive prospect for explaining fast radio bursts~\cite{barrau2014b,rovelli2017}. However, with such a severe constraint available we can see that the energy in radio is $\lesssim 10^{13}$ erg. Alternatively we can assume the Planck star supplies a large portion of its energy to radio emission and thus can explain a fast radio burst. Then we must conclude that it is $f_{pbh}$ that is extremely tiny and then that very few radio bursts could be the result of planck star events.

Additionally, we repeated this calculation when $\mu = 10^{-9}$ M$_{\odot}$ and $\sigma = 0.5$, shown in \cite{kuhnel2017} to be able to constitute $f_{pbh} \sim 1$ by evading any constraints, we still find that $\chi_0 f_{pbh} < 10^{-35}$.
 
\vspace{3cm}

\section{Conclusions}
\label{sec:conc}
We have demonstrated that, using EBL data, we can restrict the total energy emitted at low frequencies by a planck star object exploding in the present epoch to $\lesssim 10^{13}$ erg, or that only a tiny PBH population is compatible with the planck star hypothesis. This constraint depends slightly upon the choice of primordial black hole mass distribution. However, we employ a mass distribution model that is shown to approximately describe a wide range of models where the PBH over-densities are seeded by inflation~\cite{inf1,inf2,inf3}. Despite our two featured mass-function models differing in their mean PBH mass by many orders of magnitude only one order of magnitude separates their limits on the planck star low frequency emissions. We note that, although the spectral energy distribution of the explosion is unknown at low energies, we employ a simple thermal model due to its steeply peaked nature and thus do not expect the exact nature of the distribution to strongly effect these limits (as the distribution is expected to be steeply peaked anyway~\cite{barrau2014b}). Thus, these constraints cast doubt upon whether or not exploding planck stars described by the model employed here could account for fast radio bursts as speculated by \cite{barrau2014b,rovelli2017}, as their low-energy output is shown here to be comparatively small in order for emissions over all epochs to avoid exceeding the known low frequency isotropic backgrounds with both viable PBH distributions used. In addition, even if their energy production is sufficient for a fast radio burst, this would place extremely strong limits on the population of PBHs compatible with the planck star scenario. This, in turn, could imply such explosion events must be very uncommon and could explain very few fast radio burst events.  

\ack 
This work is based on the research supported by the South African
Research Chairs Initiative of the Department of Science and Technology
and National Research Foundation of South Africa (Grant No 77948).

\section*{References}
\bibliographystyle{iopart-num}
\bibliography{planck_stars.bib}

\end{document}